\documentclass[aps,prd,nofootinbib,superscriptaddress,onecolumn]{revtex4-2}

\usepackage{amsmath,amssymb,amsthm,amstext}
\usepackage{tabularx}
\usepackage{natbib}
\usepackage{graphicx}
\usepackage{color}
\usepackage{array, enumerate}
\usepackage{bm}
\usepackage{multirow}
\usepackage[breaklinks,colorlinks,citecolor=blue]{hyperref}
\usepackage{braket}
\usepackage{txfonts}
\usepackage{makecell}
\usepackage{siunitx}

\def\be{\begin{equation}}
\def\ee{\end{equation}}
\def\d{{\rm d}}
\def\pbh{\text{\tiny PBH}}
\def\abh{\text{\tiny ABH}}

\begin{document}

\title{Implications of GWTC-3 on primordial black holes from vacuum bubbles}

\author{Jibin He}
\email{hejibin20@mails.ucas.ac.cn}
\affiliation{International Centre for Theoretical Physics Asia-Pacific, University of Chinese Academy of Sciences, 100190 Beijing, China}
\affiliation{Taiji Laboratory for Gravitational Wave Universe (Beijing/Hangzhou), University of Chinese Academy of Sciences, 100049 Beijing, China}
\affiliation{School of Physics, University of Chinese Academy of Sciences, Beijing 100049, China}

\author{Heling Deng}
\email{denghe@oregonstate.edu}
\affiliation{Department of Physics, Arizona State University, Tempe, AZ 85287, USA}
\affiliation{Department of Physics, Oregon State University, Corvallis, OR 97331, USA}

\author{Yun-Song Piao}
\email{yspiao@ucas.ac.cn}
\affiliation{School of Physics, University of Chinese Academy of Sciences, Beijing 100049, China}
\affiliation{International Centre for Theoretical Physics Asia-Pacific, University of Chinese Academy of Sciences, 100190 Beijing, China}
\affiliation{School of Fundamental Physics and Mathematical Sciences, Hangzhou Institute for Advanced Study, UCAS, Hangzhou 310024, China}
\affiliation{Institute of Theoretical Physics, Chinese Academy of Sciences, P.O. Box 2735, Beijing 100190, China}

\author{Jun Zhang}
\email{zhangjun@ucas.ac.cn}
\affiliation{International Centre for Theoretical Physics Asia-Pacific, University of Chinese Academy of Sciences, 100190 Beijing, China}
\affiliation{Taiji Laboratory for Gravitational Wave Universe (Beijing/Hangzhou), University of Chinese Academy of Sciences, 100049 Beijing, China}

\begin{abstract}
\noindent
The population of black holes inferred from the detection of gravitational waves by the LIGO-Virgo-KAGRA Collaboration has revealed interesting features in the properties of black holes in the Universe. We analyze the GWTC-3 dataset assuming the detected black holes in each event had an either  astrophysical or primordial origin. In particular, we consider astrophysical black holes described by the fiducial \textsc{Power Law + Peak} distribution and primordial black holes whose mass function obeys a broken power law. These primordial black holes can be generated by vacuum bubbles that nucleate during inflation. We find that astrophysical black holes dominate the events with mass less than $\sim 30M_\odot$, whereas primordial black holes are responsible for the massive end, and also for the peak at $\sim 30M_\odot$ in the mass distribution. More than half of the observed events could come from primordial black hole mergers. We also discuss the implications on the primordial black hole formation mechanism and the underlying inflationary model.
\end{abstract}

\maketitle

\section{Introduction}

The LIGO-Virgo-KAGRA (LVK) Collaboration has so far reported about 70 events that could be confidently identified as gravitational waves emitted from black hole binary (BHB) mergers~\cite{LIGOScientific:2018mvr,Abbott:2020niy,LIGOScientific:2021djp}. 
These events and their properties are collected in the cumulative Gravitational-Wave Transient Catalog 3 (GWTC-3)~\cite{LIGOScientific:2021djp}. 
While population analyses performed on GWTC-3 show interesting substructures in the mass distribution of the detected black holes~\cite{LIGOScientific:2020kqk,LIGOScientific:2021psn}, implying more than one channel of BHB formation~\cite{Farah:2023vsc}, the physical origin of these black holes is still a topic of discussion.

It is known that a massive star can collapse to produce an astrophysical black hole (ABH), and binaries of ABHs formed by the isolation evolution of massive-star binaries and by the dynamical assembly in a dense stellar environment \cite{Mandel:2018hfr,Mapelli:2021taw} could merge at low redshift, providing plausible origins for the BHBs in GWTC-3. 
However, ABHs are not expected to form with mass in a range of $50\text{-}130 M_\odot$, a mass gap due to the pair instability~\cite{Woosley:2016hmi}. Thus they cannot explain the observed high-mass black holes without the aid of additional mechanisms. Phenomenologically, the observed black holes can be described by the \textsc{Power Law + Peak} model, where the lower bound of the power gives a peak at $\sim 10 M_\odot$ and the Gussian peak gives a peak at $\sim 35 M_\odot$~\cite{LIGOScientific:2020kqk,LIGOScientific:2021psn}.

In addition to the astrophysical origin, black holes produced before the matter-dominated era, aka primordial black holes (PBHs)\cite{Carr:2020gox,Sasaki:2018dmp,Khlopov:2008qy}, could also form binaries that merge within the age of our Universe, making contributions to the LVK detections. 
The primordial origin was suggested in Refs.~\cite{Bird:2016dcv, Clesse:2016vqa, Sasaki:2016jop} soon after the first BHB was observed, and was further investigated with updated data in Refs.~\cite{Hutsi:2020sol,DeLuca:2021wjr,Franciolini:2021tla}. 
Unlike ABHs, PBHs can in principle have mass ranging from the Plank mass $(\sim10^{-38} M_\odot)$ to orders of magnitude larger than the solar mass, subject to the abundance constraints from $\gamma$ ray, microlensing, cosmic microwave background, and many other observations at corresponding mass bands (see Ref.~\cite{Carr:2020gox} and references therein). 
The mass distribution of PBHs depends on the formation mechanisms and has been used to distinguish between early universe scenarios recently~\cite{Cai:2023ptf}. 
In the most widely studied category of PBH mechanism, perturbative quantum fluctuations during inflation can give rise to large overdensities that could collapse into black holes during the postinflation evolution.\footnote{There are some recent discussions regarding whether single-field inflation allows the formation of (large) PBHs~\cite{Kristiano:2022maq,Inomata:2022yte,Riotto:2023hoz,Choudhury:2023vuj,Kristiano:2023scm}.} 
PBHs generated in this way approximately obey a (skew-)lognormal mass distribution \cite{Dolgov:1992pu,Kannike:2017bxn,Gow:2020cou}. The perturbative-quantum-fluctuation-led PBHs and the associated lognormal mass distribution are typically assumed when investigating the primordial origin of the GWTC-3 black holes~\cite{Hutsi:2020sol,DeLuca:2021wjr,Franciolini:2021tla}.

It is noticed in Ref.~\cite{Garriga:2015fdk} that nonperturbative quantum effects could also play a role in PBH formation. 
In particular, spherical domain walls and vacuum bubbles could nucleate during inflation via quantum tunneling in a multifield potential~\cite{Basu:1991ig,Coleman:1980aw}.\footnote{A different formation channel of vacuum bubbles can be found in Ref.~\cite{Maeso:2021xvl}.} 
For a sufficiently small nucleation rate, the nucleated walls and bubbles will expand during inflation, typically without interacting with each other. After inflation, the walls and bubbles will start receding relative to the Hubble flow at some point, (with some of them) eventually forming PBHs. The resulting PBHs are referred to as subcritical or supercritical depending on whether their mass is below or beyond the critical mass parameter $m_*$, which is determined by the specific underlying physical model.
While a subcritical PBH is a typical black hole, a space-time singularity enclosed by the black hole horizon given general relativity, a supercritical black hole also contains a space-time patch that evolves independently with generally nonsingular future infinities, which in other words is a baby universe. 
This scenario has been further investigated in Refs.~\cite{Deng:2016vzb,Deng:2017uwc,Deng:2020mds,Deng:2021ezy,Wang:2018cum}. 
It is understood that PBHs forming from domain walls and vacuum bubbles typically have a broken power law mass distribution with a break mass at the critical mass $m_*$.

In this work, we shall discuss the possibility that, in addition to the fiducial \textsc{Power Law + Peak} ABHs, part of the GWTC-3 black holes are quantum-tunneling-led PBHs, and shall use GWTC-3 to constrain the broken power law PBH mass distribution. The purpose is twofold: (i) to examine if PBHs forming from domain walls/vacuum bubbles can account for some features observed in the GWTC-3 population analyses; and (ii) to investigate the implications of GWTC-3 on the physics that leads to the formation of primordial domain walls and vacuum bubbles. 

The rest of the paper is organized as follows. In Sec. \ref{sec:merger} we review the mass functions and BHB merger rates in PBH and ABH models considered in this work. The standard hierarchical Bayesian inference method used to analyze the GWTC-3 is reviewed in Sec. \ref{sec:bayes}. In Sec. \ref{sec:results} we summarize the results of analyzing the GWTC-3 dataset. Sec. \ref{sec:PBH} is devoted to a brief review of our PBH mechanism and the implications of our results on the mechanism. Conclusions are summarized and discussed in Sec. \ref{sec:conclusions}.

\section{Black Hole Merger Rates and Mass Distributions}\label{sec:merger}

We begin with PBHs, the merger rate of which in the early universe has been extensively studied in the literature \cite{Sasaki:2016jop,Raidal:2017mfl,Ali-Haimoud:2017rtz,Raidal:2018bbj}. In this work, we do not consider the mass growth of PBHs caused by accretion, nor the spin distribution of PBHs. The former is highly model dependent \cite{DeLuca:2020bjf}, while the spin features in GWTC-3 are not very informative. In this case, the differential merger rate for binary black holes of masses $m_1$ and $m_2$ is given by \cite{Raidal:2018bbj}

\begin{align}
	\frac{\d R_\pbh }{\d m_1 \d m_2}
	& = 
	\frac{1.6 \times 10^6}{{\rm Gpc^3 \, yr}} 
	f_{\pbh}^{\frac{53}{37}} \,
	\eta^{-\frac{34}{37}} 
	\left ( \frac{M}{M_\odot} \right )^{-\frac{32}{37}}  
	\left ( \frac{t}{t_0} \right )^{-\frac{34}{37}}  
	{\cal S }\left ( M, f_{\pbh},\psi_\pbh  \right )
	\, \psi_\pbh (m_1)\, \psi_\pbh (m_2)\, ,
	\label{PBHrate}
\end{align}
where $\eta = m_1 m_2/(m_1+m_2)^2$, $M= m_1+m_2$, $f_\pbh$ is the fraction of PBHs in dark matter, $t$ is the time when the merger occurs, $t_0$ is the current age of the Universe, and ${\cal S} \equiv {\cal S}_1 \times {\cal S}_2$ is the suppression factor accounting for the possible disruption of binaries due to the surrounding environments. 
The first term can be estimated as
\begin{align}
{\cal S}_1 (M, f_\pbh, \psi) \approx 1.42 \left[\frac{\langle m^2  \rangle / \langle m  \rangle ^2}{\bar{N}(y)+C} + \frac{\sigma_M^2}{f^2_\pbh}\right]^{-21/74} \exp\left[-\bar{N}(y)\right]\,,
\end{align}
where $m$ is the PBH mass, $\bar{N}(y) \simeq M f_\pbh / [\langle m \rangle  (f_\pbh+\sigma_M)]$ is the expected number of PBHs within a comoving sphere of radius $y$ around the initial PBH pair, $\sigma_M \simeq 0.004$  is the rescale variance of matter density perturbations at the time of binary formation, and $C(f_\pbh)$ is a fitting function given in Ref.~\cite{Hutsi:2020sol}. 
The second term is 
\begin{align}
{\cal S}_2 \approx {\rm min} [1,9.6 \times 10^{-3}x^{-0.65}{\rm exp} (0.03{\rm ln}^2 x)]
\end{align}
where $x \equiv (t(z)/t_0)^{0.44} f_\pbh$. In the case of small PBH abundance ($f_\pbh \sim 0.001$), ${\cal S}$ can be estimated as ${\cal S}_1$. Finally, $\psi_\pbh(m)$ is the PBH mass distribution, which is defined by
\begin{equation}
\psi_\pbh(m) \equiv \frac{m}{\rho_\pbh}\frac{\text{d}n_\pbh}{\text{d}m},
\end{equation}
where $\d n_\pbh$ is the PBH number density within the mass range $(m,m+\text{d}m)$, and $\rho_\pbh$ is the PBH energy density. The mass distribution is normalized such that $\int \psi_\pbh \text{d}m=1$. Motivated by the formation mechanism of PBHs from domain walls/vacuum bubbles, we are interested in a PBH mass function described by a broken power law~\cite{Garriga:2015fdk,Deng:2016vzb,Deng:2017uwc,Deng:2020mds,Deng:2021ezy}
\begin{equation}\label{psipbh}
	\psi_{\pbh}\left(m|m_*, \alpha_1, \alpha_2\right)=\frac{1}{m_{*}\left(\alpha_{1}^{-1}-\alpha_{2}^{-1}\right)}
	\begin{cases}
		(m/m_{*})^{\alpha_{1}-1}, & m<m_{*}\\
		(m/m_{*})^{\alpha_{2}-1}, & m>m_{*}
	\end{cases}\, 
\end{equation}
where $m_*$ is the critical mass with $\alpha_1$ and $\alpha_2$ being the spectral indices for the subcritical and supercritical PBHs respectively. It is also useful to introduce 
\begin{equation}
f(m)\equiv m f_\pbh \psi_\pbh(m)
\end{equation}
as the fraction of dark matter in PBHs at $m$ within the mass range $\Delta m \sim m$.

In our analysis, we also consider the astrophysical origin of BHBs. Following Refs.~\cite{DeLuca:2021wjr, LIGOScientific:2018jsj}, the differential merger rate of ABHs can be written as 
\begin{align}\label{rateABH}
	\frac{\d R_\abh}{\d m_1\d m_2 } =  {\cal N} ~\bar{R}_\abh ~ (1+z)^\kappa~\pi(m_1, m_2)
\end{align}
where $\bar{R}_\abh$ is the local merger rate at redshift $z=0$, ${\cal N}$ is a normalization factor ensuring $R_\abh(z=0)=\bar{R}_\abh$, and $\kappa \simeq 2.7$ describes the merge rate evolution with redshift~\cite{Madau:2014bja, LIGOScientific:2021psn}. $\pi(m_1, m_2)$ depends on the mass distribution of ABHs. In the literature, there are different proposals for the ABH mass distribution. In this work, we shall consider the \textsc{Power Law + Peak} model~\cite{LIGOScientific:2020kqk,Talbot:2018cva}, whose primary mass distribution obeys
\begin{align}\label{psiabhp}
	\psi_\abh(m_1 | \lambda_{\rm peak}, m_{\rm min}, m_{\rm max},\zeta,\mu_m,\sigma_m ,\delta_m) \propto \left[(1-\lambda_{\rm peak}){\cal B}(m_1|-\zeta,m_{\rm max})+\lambda_{\rm peak}{\cal G}(m_1|\mu_m,\sigma_m)\right] S(m_1|m_{\rm min},\delta_m),
\end{align}
where ${\cal B}$ is a normalized power law distribution with a spectral index of $-\zeta$, and ${\cal G}$ is a normalized Gaussian distribution with  mean  $\mu_m$ and  width  $\sigma_m$. $\lambda_{\rm peak}$ is the fraction of the Gaussian component in the primary mass distribution.  $S(m_1,m_{\rm min}, \delta_m)$ is a smoothing function which rises from 0 to 1 over the interval $(m_{\rm min}, m_{\rm min}+\delta_m)$,
\begin{equation}
	\label{eq:smoothing}
	S(m \mid m_{\rm min}, \delta_m) = \begin{cases}
		0, & m< m_{\rm min} \\
		\left[f(m - m_{\rm min}, \delta_m) + 1\right]^{-1}, & m_{\rm min} \leq m < m_{\rm min}+\delta_m \\
		1, & m\geq m_{\rm min} + \delta_m
	\end{cases}
\end{equation}
with
\begin{equation}
	f(m', \delta_m) = \exp \left(\frac{\delta_m}{m'} + \frac{\delta_m }{m' - \delta_m}\right).
\end{equation}
For the \textsc{Power Law + Peak} ABH model, the conditional mass ratio distribution satisfies
\begin{align}
	 \pi_\abh(m_1, m_2) \propto C(m_1) \psi_\abh(m_1 | \lambda_{\rm peak}, m_{\rm min}, m_{\rm max},\zeta,\mu_m,\sigma_m ,\delta_m)\,q^{\beta_q},
\end{align}
where $q=m_2/m_1$ and $C(m_1)$ is a normalization factor.

In the later analysis, we shall consider two hypotheses: (1) All BHBs are of astrophysical origin; (2) BHBs could be either astrophysical or primordial. In the former hypothesis, the merger rate is simply given by Eq.~\eqref{rateABH}, i.e., $\d R/\d m_1 \d m_2 =\d R_\abh/\d m_1 \d m_2$. In the latter hypothesis, which we shall refer to as the ABH-PBH model, the mass distribution of ABHs is described by the \textsc{Power Law + Peak} model \eqref{psiabhp} while the mass distribution of PBH is given by Eq.~\eqref{psipbh}, and the total merger rate is given by
\begin{align}
\frac{\d R}{\d m_1\d m_2 } = \frac{\d R_\abh}{\d m_1\d m_2 } + \frac{\d R_\pbh}{\d m_1\d m_2 }\, .
\end{align} 

\section{Hierarchical Bayesian Inference}\label{sec:bayes}

The hierarchical Bayesian analysis~\cite{Mandel:2018mve,Vitale_2021} is currently extensively utilized in the analysis of the BHB population~\cite{Wong:2020yig,DeLuca:2021wjr,Zheng:2022wqo,Franciolini:2022tfm}. In this section, we will provide a brief introduction to our parameter analysis.  For each population model, the \textsc{Power Law + Peak} ABH model, or the ABH-PBH model, we marginalize over the parameters of individual events to find the posterior distributions of the model parameters. To be concrete, we label the parameters of individual events, i.e., the intrinsic parameters, as $\theta$, and the parameters of population models, i.e., the hyperparameters, as $\Lambda$. In practice, we consider $\theta = \{m_1,\, m_2,\, z\}$. We have $\Lambda =\{\bar{R}_\abh, \,\lambda_{\rm peak},\,  m_{\rm min},\, m_{\rm max},\, \zeta ,\,\mu_m, \,\sigma_ m,\, \delta_m,\, \beta_q\}$ for the \textsc{Power Law + Peak} ABH model. For the ABH-PBH model, the hyperparameters also include $\{f_\pbh,\,m_*,\, \alpha_1,\, \alpha_2\}$ besides the ones in the ABH model. The hyperparameters of the models and their priors $\pi(\Lambda)$ used in the hierarchical Bayesian inference are listed in Table~\ref{tab:para}.

Given a population model of parameters $\Lambda$, the likelihood of a dataset ${\bm d}$ is
\begin{align}\label{likefull}
{\cal L}({\bm d} | \Lambda) \propto e^{-N(\Lambda)\xi(\Lambda)} [N(\Lambda)]^{N_{\rm det}} \prod_{i=1}^{N_{\rm det}} \int {\cal L}(d_i|\theta)\, \pi(\theta|\Lambda)\,d\theta \, .
\end{align}
Here ${\bm d} = \{d_i\}$ with $i$ labeling an individual event from the considered detections. $N_{\rm det}$ is the number of detected merger events considered in the analysis. $N(\Lambda)$ is the total number of merging events expected by the model, and hence depends on the population model. Given the differential merger rate, the differential expected number of events can be evaluated as~\cite{LIGOScientific:2018jsj}
\begin{align}
\frac{\d N}{\d m_1\d m_2 \d z} = T_{\rm obs} \frac{1}{1+z} \frac{\d V_c}{\d z} \frac{\d R}{\d m_1 \d m_2} 
\end{align}
where $T_{\rm obs}$ is the effective observing time, $(1+z)^{-1}$ accounts for the time redshift at the source frame, and $V_c$ is the comoving volume. In particular, we assume a flat $\Lambda$CDM universe in which
\begin{align}
\frac{\d V_c}{ \d z} = \frac{4\pi}{H_0} \frac{D_c^2(z)}{E(z)}\, ,
\end{align}
with $E(z) = \sqrt{\Omega_M(1+z)^3+\Omega_\Lambda}$ and the comoving distance,
\begin{align}
D_c(z) = \frac{1}{H_0} \int_0^z \frac{\d z'}{E(z')}.
\end{align}
The comoving distance $D_c(z)$ relates to the luminosity distance $D_L(z)$ by $D_c(z) = D_L(z)/(1+z)$. In our calculation, we take $H_0 = 67.9 {\rm km/s/Mpc}$ and $\Omega_M = 0.3065$ from the Planck 2015 results~\cite{Planck:2015fie}. Assuming a log-uniform prior on the total expectation number $N$, we can marginalize Eq.~\eqref{likefull}. By doing so, we obtain 
\begin{align}\label{likeint}
{\cal L}({\bm d} | \Lambda) \propto  \prod_{i=1}^{N_{\rm det}} \frac{\int {\cal L}(d_i|\theta) \pi(\theta|\Lambda) d\theta}{\xi(\Lambda)}\, .
\end{align}
Finally, $\xi(\Lambda)$ in Eq.~\eqref{likefull} is the detection fraction, i.e., the fraction of binaries that we expect to detect given the model with hyperparameter $\Lambda$. Formally,
\begin{align}
	\xi (\Lambda) = \int p_{\rm det}(\theta)\,\pi(\theta|\Lambda)\d \theta \, ,
\end{align}
where $p_{\rm det}(\theta)$ is the detection probability of an event with parameters $\theta$, and $\pi(\theta|\Lambda)$ is the prior of $\theta$ given the population model of parameters $\Lambda$. We utilize simulated signals of injections provided by LVK to estimate the detection fraction, which can be approximated as
\begin{align}
\xi(\Lambda) \propto \tilde{\xi}(\Lambda) = \frac{1}{N_{\rm inj}}\sum_{j=1}^{N_{\rm tri}} \frac{\pi(\theta_j|\Lambda)}{p_{\rm draw}(\theta_j)}
\end{align}
where $N_{\rm inj}$ is the number of injections, $N_{\rm tri}$ is the number of the injections which can be detected, and $p_{\rm draw}(\theta_j)$ is the distribution from which the injections are drawn~\cite{LIGOScientific:2018jsj,LIGOScientific:2020kqk}.
In practice, we replace the integrals in Eq.~\eqref{likeint} with weighted averages over discrete samples,
\begin{align}\label{likesum}
{\cal L}({\bm d} | \Lambda) \propto \prod_{i=1}^{N_{\rm det}} \frac{1}{\xi(\Lambda)} \frac{1}{n_i}\sum_{j=1}^{n_i}\frac{\pi(\theta_{ij}|\Lambda)}{\pi(\theta_{ij})} \, ,
\end{align}
where $\theta_{ij}$ denotes the intrinsic parameters of the $j$th sample of the $i$th event, and $\pi(\theta_{ij})$ is the prior on the binary parameters used when performing the parameter estimation. The posterior of the hyperparameters $\Lambda$ given the observed dataset ${\bm d}$, $p(\Lambda|{\bm d})\propto {\cal L}({\bm d} | \Lambda)\pi(\Lambda)$, is obtained by {\footnotesize EMCEE}~\cite{2013PASP} and {\footnotesize DYNESTY}~\cite{Speagle_2020}.

The dataset ${\bm d}$ we consider consists of 69 BHB merger events (listed in Table 28 of Ref.~\cite{LIGOScientific:2021psn}). These events have a false alarm rate ${\rm FAR}<1 yr^{-1}$. To avoid the potential impact of neutron star coalescences in our analysis, the dataset does not include any events with  masses less than $3 M_\odot$. The posterior sample data for these events was published by LVK~\cite{ligo_scientific_collaboration_and_virgo_2021_5546663}, and we utilize the  CO1:Mixed samples. Most of the selected events involve black holes with masses below $50 M_\odot$, but there are some exceptions, such as GW190521, which contains black holes with masses exceeding $50 M_\odot$ and therefore falls within the pair-instability mass gap.

\setlength{\extrarowheight}{1.5pt}
\begin{table*}
	 \centering
	\begin{ruledtabular}
		\begin{tabular} {clrrl}
			Parameter &Prior   &ABH-PBH & ABH & Description \\ 
			\hline
			\ & \ &  \  & \  &\textsc{Broken Power Law PBH} \\
			$m_*/M_\odot$ & $[5,50]$ &   $ 29.64_{-1.82}^{+2.01} $   &\ &The critical mass \\
			$\log_{10}f_{\pbh}$ & $[-5,-1]$  &  $-3.09_{-0.08}^{+0.06}$   &\ & Logarithmic fraction of dark matter  in PBHs \\
			$\alpha_1$ & $[1,15]$  &  $10.23_{-3.74}^{+3.23}$ &  \ &Spectral index of the mass function of subcritical PBHs\\
			$\alpha_2$ & $[-15,-1]$ &  $-3.84_{-3.23}^{+0.89}$   & \ &Spectral index of the mass function of supercritical PBHs\\

                 \hline 
			\ & \ &     \ &   \ &\textsc{Power Law + Peak ABH} \\
			$\bar{R}_{\abh}/{\rm Gpc^{-3}yr^{-1}} $ & $[5,50]$  & $17.23_{-4.06}^{+4.82}$  &  $17.66_{-3.75}^{+4.80}$  & Integrated merger rate of ABHs at $z = 0$  \\
 			$\log_{10} \lambda_{\rm peak} $ & $[-6,0]$ & $-3.23_{-1.86}^{+1.23}$  & $-1.70_{-0.36}^{+0.29}$ & Fraction of the Gaussian component in the primary mass distribution \\
			$m_{\rm min}/M_\odot$  &  $[2,10]$ & $5.02_{-0.89}^{+0.66}$  &  $4.70_{-1.06}^{+0.84}$ & Minimum mass of the power law component in the primary mass distribution \\
			$m_{\rm max}/M_\odot$ &  $[30,100]$ & $72.37_{-32.63}^{+20.13}$  &$87.55_{-9.81}^{+8.64}$ & Maximum mass of the power law component in the primary mass distribution \\
			$\zeta $ & $[-4,12]$ & $3.97_{-0.79}^{+2.09}$  &$3.59_{-0.34}^{+0.37}$  & Slope of the primary mass distribution for the power law component\\
			
			$\mu_m/M_\odot$ & $[20,50]$ & $29.64_{-7.16}^{+12.10}$  &$34.03_{-1.78}^{+1.49}$  & Mean of the Gaussian component \\
			$\sigma_m/M_\odot$ & $[1,10]$ & $4.86_{-2.71}^{+3.38}$   &$3.21_{-1.44}^{+2.51}$  & Width of the Gaussian component \\
			$\delta_m/M_\odot$ & $[0,10]$ & $6.26_{-2.19}^{+2.15}$  &$5.84_{-2.09}^{+2.19}$ &Range of mass tapering on the lower end of the mass distribution\\
			$\beta_q$ &  $[-4,7]$ & $-1.57_{-1.61}^{+1.87}$   &$ 1.57_{-1.07}^{+1.49} $ & Spectral index for the power law of the mass ratio distribution \\
		\end{tabular}
	\end{ruledtabular}
	\caption{\label{tab:para}Prior and posterior credible intervals ($68\%$) of the hyperparameters of the ABH-PBH model and the  \textsc{Power Law + Peak} ABH model.}\label{tab:para}
\end{table*}

\section{Results}\label{sec:results}

\begin{figure}[t!]
	\centering
	\includegraphics[width=0.88 \linewidth]{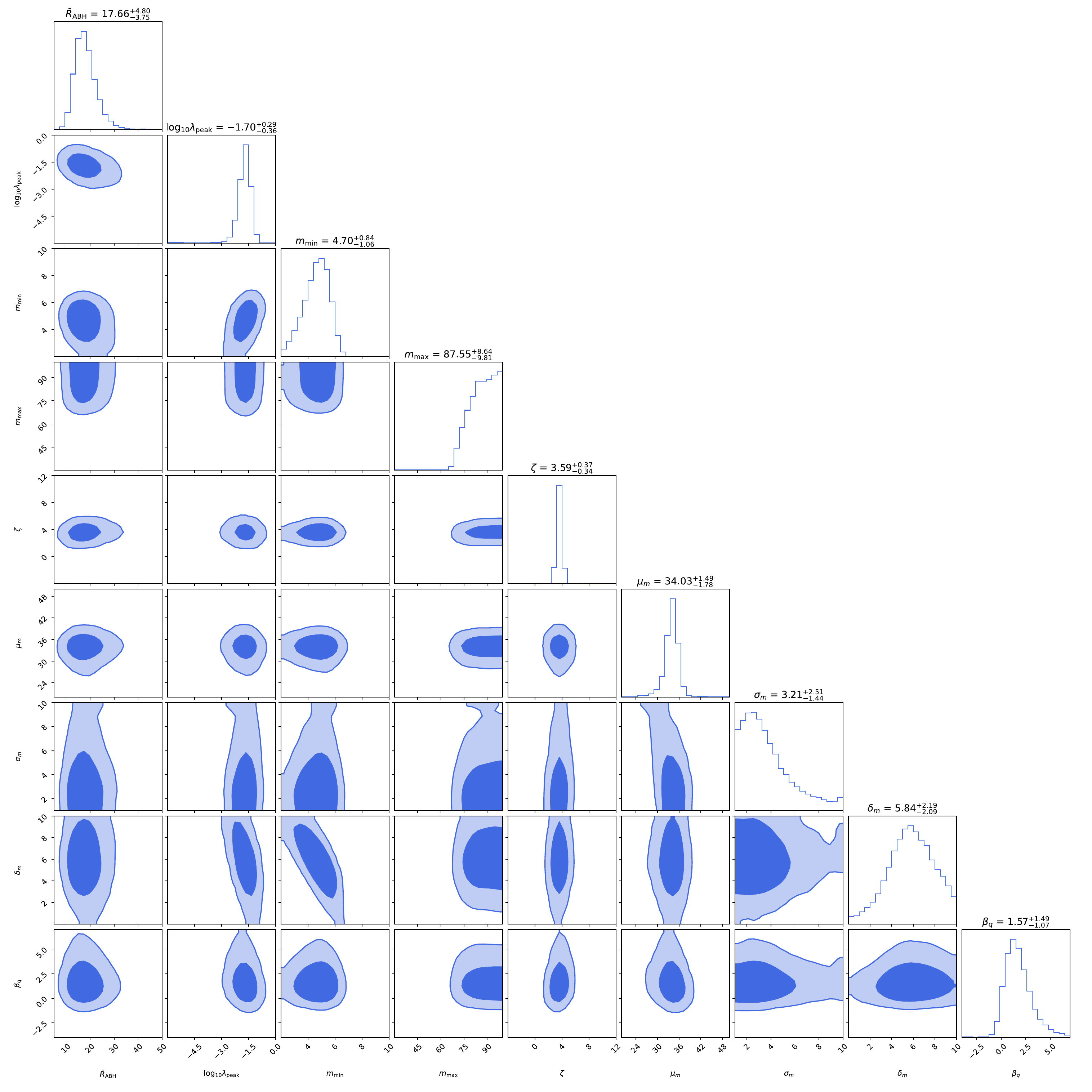}
	\caption{Posterior distributions of the hyperparameters in the \textsc{Power Law + Peak} ABH model. The values displayed at the top of the plots represent the $68\%$ credible intervals. }
	\label{fig:postPeak}
\end{figure}

\begin{figure}[t!]
	\centering
	\includegraphics[width=0.88 \linewidth]{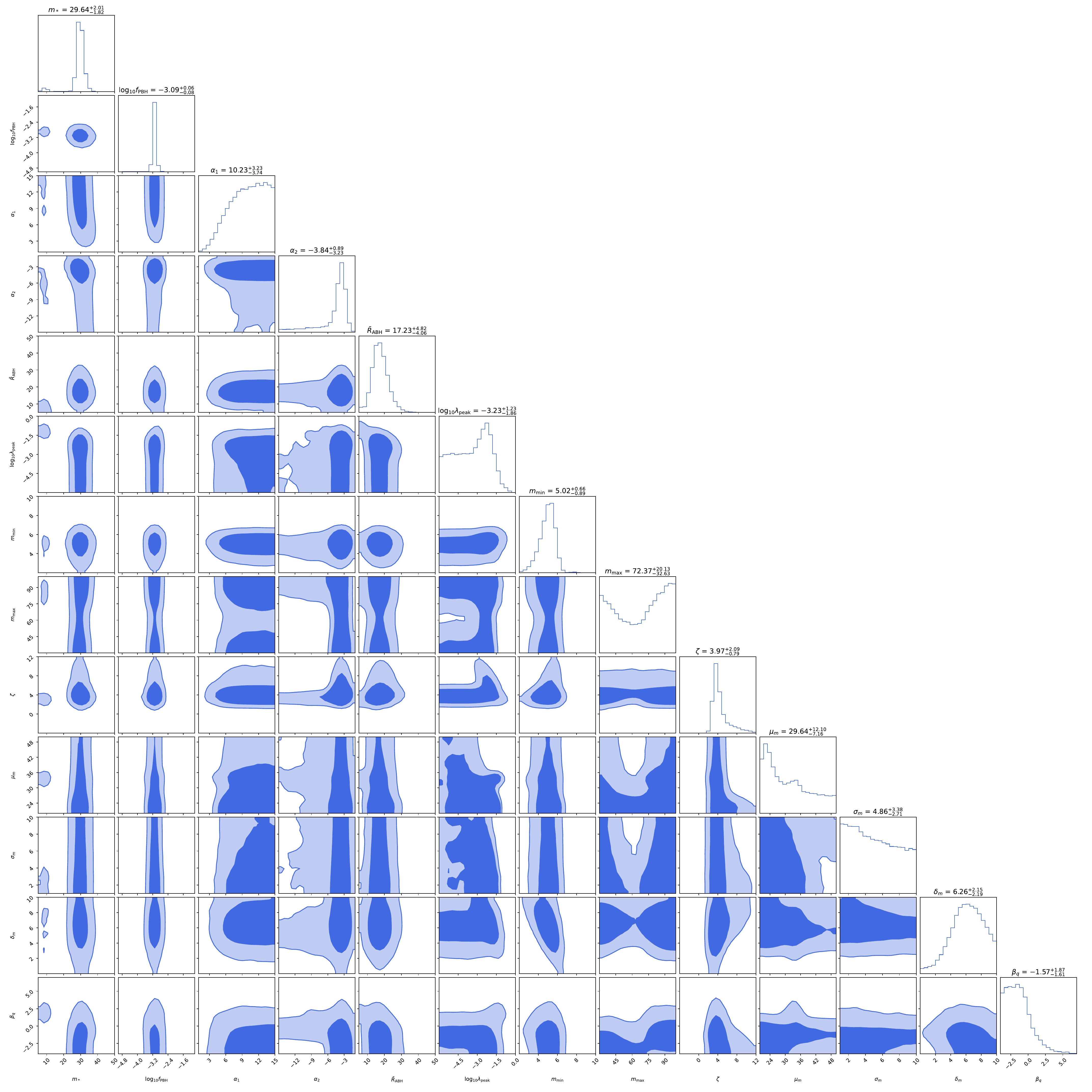}
	\caption{Posterior distributions of the hyperparameters in the ABH-PBH model. The values displayed at the top of the plots represent the $68\%$ credible intervals.}
	\label{fig:postAP}
\end{figure}

\begin{figure}[h!] 
	\centering
	\includegraphics[width=0.488 \linewidth]{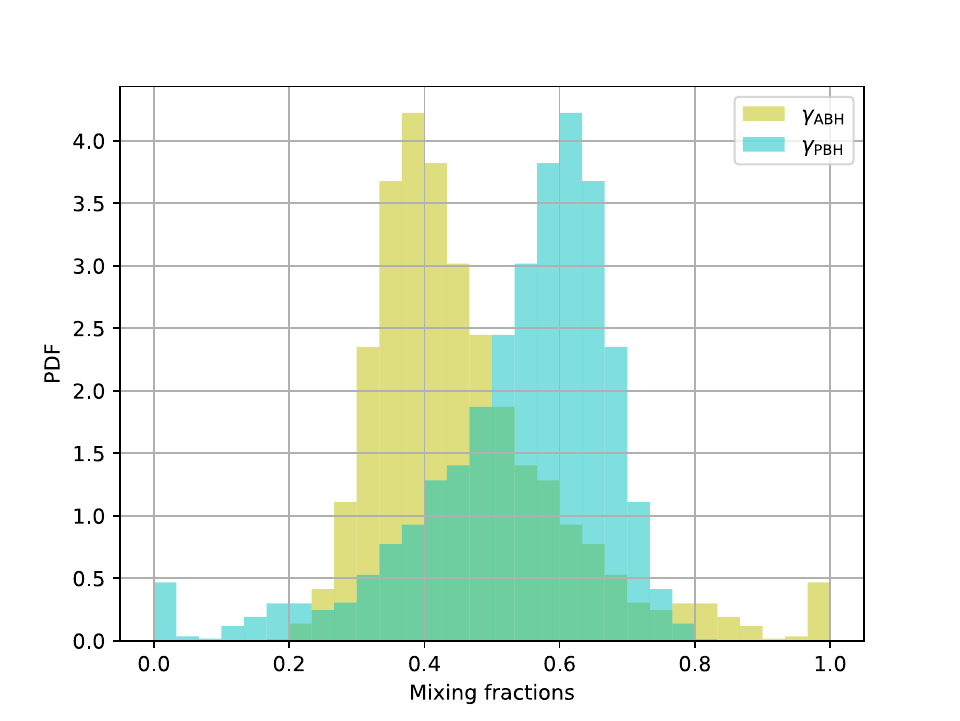}
	\caption{The posterior predictive fractions of ABHs and PBHs in GWTC-3.}
	\label{fig:hist}
\end{figure}

\begin{figure}[t!] 
	\centering
	\includegraphics[width=0.488 \linewidth]{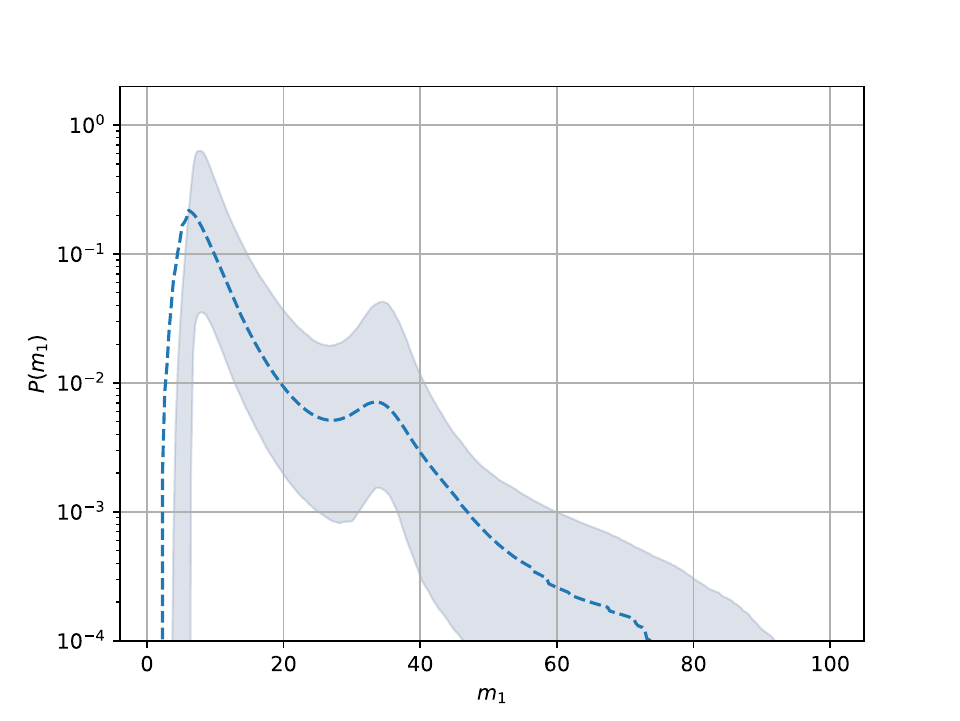}
	\caption{Posterior predictive distribution of the primary mass $m_1$ in the \textsc{Power Law + Peak}  ABH model. The dashed line represents the weighted average and the shaded area represents a 90\% credible interval. The dashed lines can be outside of the shaded areas due to the skewed posterior distribution of $\Lambda$.}
	\label{fig:massA}
\end{figure}

The posterior credible intervals ($68\%$) of the hyperparameters of the  \textsc{Power Law + Peak} ABH model and the ABH-PBH model  are listed in Table~\ref{tab:para} (also see  Fig.~\ref{fig:postPeak} and Fig.~\ref{fig:postAP} for the posterior distributions). For the ABH model, our values are largely consistent with the  results obtained by LVK. For the ABH-PBH model, we find that the observed BHBs can be best fitted  if the PBH density is about $0.1\%$ of the dark matter density. In order to get an intuition of the fraction of PBHs in the observed BHBs, we follow Ref.~\cite{DeLuca:2021wjr} and define the proportion of PBHs and ABHs in the ABH-PBH model
\begin{align}
	\gamma_\pbh &\equiv N^\text{\tiny det}_\text{\tiny PBH}/(N^\text{\tiny det}_\text{\tiny ABH}+ N^\text{\tiny det}_\pbh)\, ,\\
	\gamma_\abh &\equiv  1-\gamma_\pbh.
\end{align}
The posterior distributions of $\gamma_\pbh$ and $\gamma_\abh$ in the ABH-PBH model are shown in Fig.~\ref{fig:hist}, according to which the PBHs could account for about 60\% of the observed BHBs. 
\begin{figure}[t!]
	\centering
	\includegraphics[width=0.488 \linewidth]{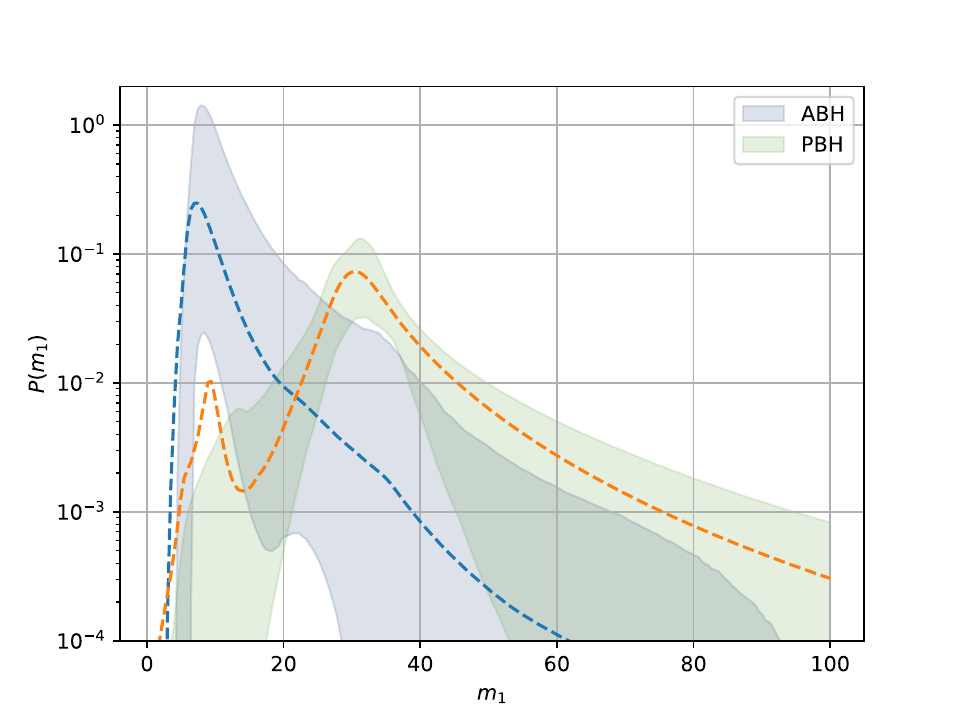}
	\includegraphics[width=0.488 \linewidth]{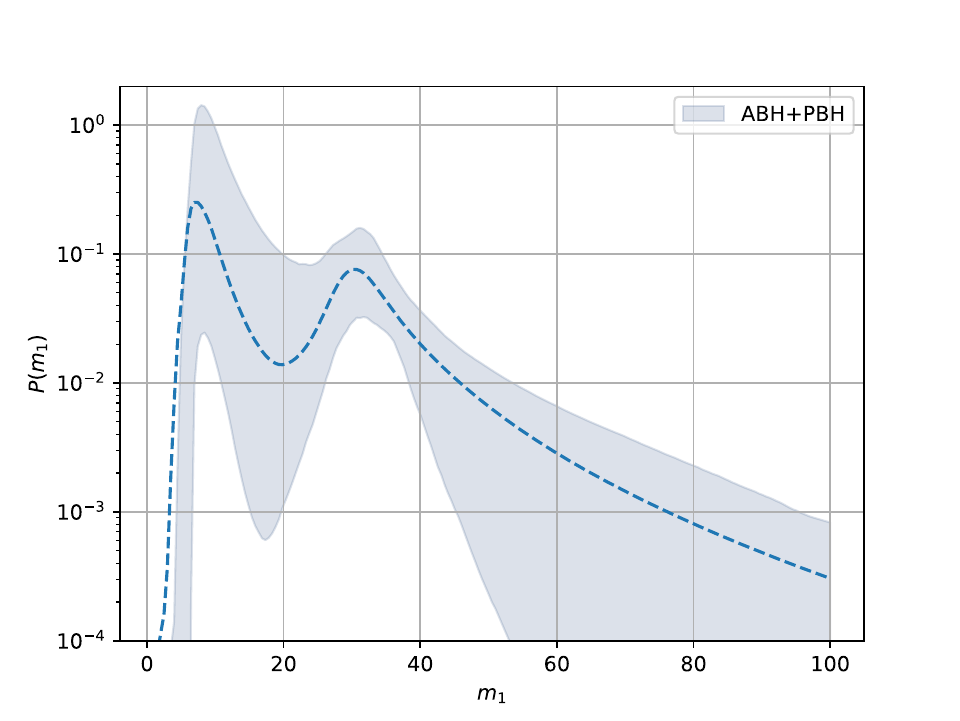}
	\caption{Posterior predictive distributions of the primary mass $m_1$ in the ABH-PBH model. The left plot shows the distribution of ABHs and PBHs separately, while the right plot shows the sum of both ABHs and PBHs. The dashed lines represent the weighted average and the shaded areas represent a 90\% credible interval. The dashed lines can be outside of the shaded areas due to the skewed posterior distribution of $\Lambda$.}
	\label{fig:mass}
\end{figure}
We also show the posterior predictive distributions of the primary mass $m_1$ of both hypotheses in Fig.~\ref{fig:mass} and Fig.~\ref{fig:massA}. For the ABH-PBH model, the weighted average of PBH has a minor peak at $\sim 8 M_\odot$, which is due to a smaller mode in the posterior distribution of $m_{*}$ at $\sim 8 M_\odot$. However, this mode is too small to have a noticeable impact on the 90\% credible interval. We find the distribution of the primary mass in the ABH-PBH model manifests two peaks, which has also been observed in the phenomenological population analysis performed in Refs.~\cite{LIGOScientific:2021psn, Farah:2023vsc}, indicating our ABH-PBH model is capable of explaining the observed data. Moreover, we find that black holes with mass less than $20 M_\odot$ as well as the peak at $\sim 10 M_\odot$ are dominated by ABHs, while black holes with mass greater than $20 M_\odot$ are more likely to be PBHs. In particular, the mass function of PBHs is likely to peak at $\sim 30 M_\odot$ with the subcritical black holes being suppressed given the large best-fit value of $\alpha_1$. We shall discuss the implications of the posterior PBH mass distribution in Sec.~\ref{sec:PBH}.

Now we consider the \textsc{Power Law + Peak} ABH model as our benchmark and calculate the Bayes factor between the ABH-PBH model and the ABH model 
\begin{equation}
	{\cal B}^{\rm ABH-PBH}_{\rm ABH} = \frac{ Z_{\rm ABH-PBH}}{Z_{\rm ABH}},
\end{equation}
where
\begin{equation}
	Z_{\cal M} \equiv \int  {\cal L}({\bm d} | \Lambda)\pi(\Lambda) \d \Lambda
\end{equation}
is the evidence for model ${\cal M}$. The evidence of each model is obtained by {\footnotesize DYNESTY}, which gives ${\cal B}^{\rm ABH-PBH}_{\rm ABH}\approx 10^{1.83}$. This indicates that our ABH-PBH model is strongly favored over the fiducial \textsc{Power Law + Peak} ABH model.

\section{PBH's from vacuum bubbles}\label{sec:PBH}

In this section, we discuss the production of PBHs whose mass function obeys a broken power law. It was proposed and investigated in a series of works that PBHs can be formed by spherical domain walls and vacuum bubbles that nucleate during inflation \cite{Garriga:2015fdk, Deng:2016vzb, Deng:2017uwc,Deng:2020mds,Deng:2021ezy}. We shall focus on the case of vacuum bubbles, introducing in more details the formation mechanism and discussing the implication of the GWTC-3 data on the model. For domain walls, the mass distribution of the resulting PBHs usually has $\alpha_2=-1/2$, which is not compatible with the $\alpha_2$ inferred from GWTC-3.

\subsection{Mechanism}
If the inflaton field lives in a multidimensional potential, it may tunnel from the quasi-de Sitter vacuum to another vacuum of a lower energy scale. 
As a result, bubbles constantly pop out in space at a certain nucleation rate, and expand at a speed close to the speed of light. 
After inflation ends, inflaton outside the bubbles rolls down to the Universe's present vacuum, decaying into hot radiation, while the rapidly expanding bubbles run into the radiation fluid. If the energy scale of the bubble interior is larger than that of our vacuum, such a bubble will eventually come to a stop and start receding with respect to the Hubble flow, because all forces acting on the bubble wall, including the vacuum pressure, the wall tension and possible friction from the radiation fluid, point inward. The fate of the bubble depends on its size. A small bubble could collapse into a black hole after it reenters the cosmological horizon. This kind of bubble and its resulting PBH are what we refer to as subcritical. For a sufficiently large bubble, the bubble wall will also start receding with respect to the Hubble flow at some point after inflation for the same reason as in the case of a subcritical bubble, but the bubble will never collapse due to the inflation occurring in the bubble interior. As a result, a wormhole forms, connecting our Universe and the bubble, and eventually pinches off, leaving a black hole in our Universe and a space-time that is causally disconnected from our Universe, i.e., a baby universe. This kind of bubble and its resulting PBH are what we refer to as supercritical.

The mass of the resulting black holes can be found by studying the evolution of the bubble, the details of which are investigated in Refs. \cite{Garriga:2015fdk, Deng:2017uwc,Deng:2020mds,Deng:2021ezy}.
Neglecting the friction from the radiation fluid, the bubble motion after inflation is determined by the following parameters: the inflationary scale $\eta_{i}$, the energy scale of the bubble interior $\eta_{\text{b}}$, the wall tension scale $\eta_{\sigma}$, and the bubble wall's Lorentz factor $\gamma$ at the end of inflation. From these parameters, along with the bubble size at the end of inflation, one finds how the bubble expands by numerically solving the bubble wall's equation of motion. The resulting black hole mass $m$ can then be estimated by the bubble size at time $t_{s}$, when the bubble comes to a stop with respect to the Hubble flow. In the subcritical regime, assuming that the bubble mass is dominated by the interior vacuum, it can be shown that black holes formed by bubble collapse have mass $m\sim\eta_{\text{b}}^{4}t_{s}^{3}/M_{\text{Pl}}^{6}$.
On the other hand, the resulting black holes from supercritical bubbles have mass estimated as $m\sim t_{s}$ \cite{Deng:2020mds,Deng:2021ezy}. Equating these two gives the critical mass that connects the two regimes:
\begin{align}
m_{*}\sim M_{\text{Pl}}^{3}/\eta_{\text{b}}^{2}.\label{m*}
\end{align}

In the above analysis, the bubble is assumed to be perfectly spherical.
However, at the time of nucleation, there are inevitable quantum fluctuations
in the bubble wall. When a subcritical bubble collapses, these fluctuations
grow and the bubble may fragment into smaller pieces, which will disintegrate
into relativistic particles. This effect certainly hinders the formation
of black holes from small subcritical bubbles. It was found in Ref. \cite{Deng:2017uwc} that in order for fluctuations not to break the shrinking bubble,
the resulting black hole should at least have mass
\begin{equation}
m_{\text{\tiny{F}}}\sim\eta_{\text{b}}^{-2}\left(\frac{\eta_{i}^{4}M_{\text{Pl}}}{\eta_{\sigma}^{3}}\right)^{3/2}.\label{wall_fluctuation}
\end{equation}
This gives a lower bound to the black hole mass in the subcritical regime. On
the other hand, supercritical bubbles are not subject to this constraint.
Therefore, if $m_{\text{\tiny{F}}}<m_{*}$, the minimum
black hole mass is given by $\sim m_{\text{\tiny{F}}}$; if $m_{\text{\tiny{F}}}>m_{*}$, then most subcritical bubbles
would not turn into black holes, and the minimum black hole
mass is given by $\sim m_{*}$.

\begin{figure}[t!]
	\centering
	\includegraphics[width=0.8 \linewidth]{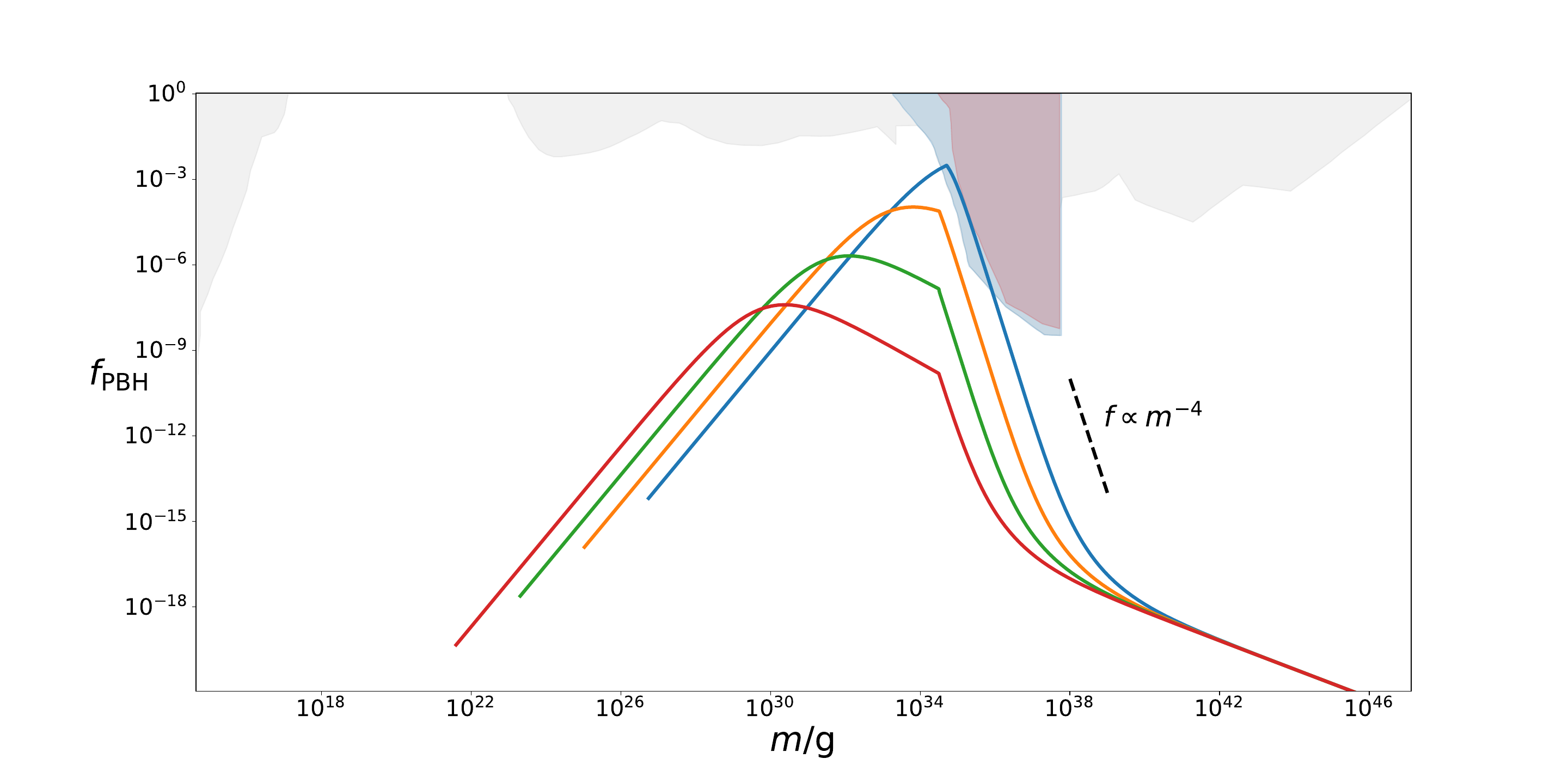}
	\caption{Observational constraints on the fraction of the dark matter in (monochromatic) PBHs $f_\pbh$ (shaded regions; adapted from Fig. 10 in Ref. \cite{Carr:2020gox}) and several examples of the PBH mass function (four curves) considered in this work. From the bottom curve (red) to the top (blue), we increase the value of the Lorentz factor $\gamma$, with all other parameters fixed.
	}
	\label{fig:fPBH}
\end{figure}

Bubbles formed at different times expand to different sizes. By working out the bubble dynamics during inflation, and assuming a constant bubble nucleation rate $\kappa$, one obtains the size distribution of the bubbles when inflation ends ($\propto \kappa$). Then by the relation of $t_s$ and $m$, we obtain the mass distribution of the black holes \cite{Deng:2016vzb}. Several examples of the mass function $f(m)$ are shown in Fig. \ref{fig:fPBH}.  We can see that $f(m)$  can be approximated by a set of broken power laws near the critical mass $m_{*}$, where there is a relatively sharp change (for example, the peak in the blue curve). The shaded regions in Fig. \ref{fig:fPBH} are observational constraints on $f_\pbh$ for monochromatic PBHs, which means all PBHs are of the same mass.\footnote{Strictly speaking, these constraints are improper for an extended mass function as our $f(m)$ \cite{Carr:2017jsz}. However, using the upper bounds of $f_\pbh$ to constrain $f(m)$ is qualitatively reasonable as long as $\psi_\pbh=f/m$ does not have a plateau over a large range.} The only window that allows PBHs to be responsible for all dark matter is restricted to $10^{17}\text{-}10^{23}\rm g$.

A noticeable feature of $f(m)$ is that PBHs in the supercritical regime near $m_*$ obey $f\propto m^\alpha$ where $\alpha\approx -4$.\footnote{By semianalytic calculations one finds $\alpha\approx -4.25$.} This is a generic result as long as the Lorentz factor $\gamma$ is sufficiently large. An assumption behind the mechanism is there is no friction exerting on the bubble wall from the radiation fluid. In the other
extreme scenario, where all fluid is reflected by the bubble wall, the resulting mass function for supercritical black holes should obey
$f\propto m^{-1/2}$. Taking mass accretion into account tends to give a shallower slope. Therefore, a power law much steeper than $m^{-4}$ is incompatible with our mechanism. If such a mass function is favored by future detection, our PBH mechanism as an explanation of the LVK events can be ruled out.

\subsection{Implications from GWTC-3}
Assuming that our ABH-PBH model is responsible for the LVK events, our analyses on the GWTC-3 dataset suggest that more than half of the LVK black holes come from PBHs. While ABHs dominate
the low-mass end in the mass distribution, larger black holes are mostly PBHs. The PBH mass function is given by
\begin{equation}
f\approx 10^{-3} m^{-4}\, \text{for \ensuremath{ m > 30M_{\odot}}},\label{PBH_result}
\end{equation}
and is suppressed at $m < 30M_{\odot}$ since $\alpha_1$ is likely to have a large value. Such a mass function can be approximated by the blue curve in Fig. \ref{fig:fPBH} for the mass range $m\gtrsim 30M_\odot\approx 6\times 10^{34}\text{g}$ (right side of the peak). Note that it looks incompatible with the light blue shaded region, which is a constraint from the nonobservation of disklike PBH accretion effects in the cosmic microwave background \cite{Serpico:2020ehh}. If the accretion is spherical instead of disklike, the light purple shaded region \cite{Serpico:2020ehh} is marginally consistent with our result.   

The mass function (\ref{PBH_result}) suggested by GWTC-3 brings several implications to our PBH mechanism:

(i) Eq. (\ref{PBH_result}) is consistent with the mass function predicted for PBHs formed from supercritical bubbles. As discussed in the previous subsection, $f\propto m^{-4}$ is a generic feature for $m \gtrsim m_*$. These PBHs can account for the GWTC-3 events at the high-mass end, as well as the peak in mass distribution at $\sim 30 M_\odot$. 

(ii) By Eq. (\ref{m*}), the critical mass $m_*$ is determined by the vacuum energy density inside the bubble. By (\ref{PBH_result}), we have
\begin{equation}
M_{\text{Pl}}^{3}/\eta_{\text{b}}^{2}\sim30M_{\odot}\to\eta_{\text{b}}\sim0.1\ \text{GeV},
\end{equation}
i.e., the energy scale of the bubble interior is $\mathcal{O}(0.1)$ GeV. 

(iii) A large $\alpha_1$ means the formation of subcritical black holes are suppressed. From the discussion in the previous subsection, this could happen if most subcritical bubbles are destroyed by wall fluctuations. By Eq. (\ref{wall_fluctuation}), we have
\begin{equation}
\eta_{\text{b}}^{-2}\left(\frac{\eta_{i}^{4}M_{\text{Pl}}}{\eta_{\sigma}^{3}}\right)^{3/2}>30M_{\odot} \to \eta_i^4>\eta^3_\sigma M_{\rm Pl}.
\end{equation}
If we further assume that the bubble wall and the bubble interior have comparable energy scales, i.e., $\eta_\sigma\sim \eta_{\rm b}$, then we have
\begin{equation}
\eta_i>10^4\ \text{GeV},
\end{equation}
which provides a lower bound to the inflationary scale.



\section{Conclusions}\label{sec:conclusions}

The analysis of GWTC-3 reported by the LVK Collaboration indicates a substructure in the mass distribution of the detected black holes. In particular, two peaks were found at $\sim10M_{\odot}$ and $\sim35M_{\odot}$ respectively, suggesting more than one channel of the formation of BHBs. The mass distribution can phenomenologically be described by the \textsc{Power Law + Peak} model, where the Gaussian peak accounts for black holes around $35M_{\odot}$. In this work, we have considered the possibility that the peak at higher masses is attributed to PBHs obeying a merger rate different from black holes on the lower mass end. In particular, we considered PBHs generated in a nonperturbative mechanism, where the black holes are formed by vacuum bubbles that nucleate during inflation. These PBHs are either ``subcritical'' or ``supercritical,'' and the mass function near the critical mass $m_{*}$ is expected to obey a broken power law. We then assumed a model where each LVK BHB is either a ABH binary from the \textsc{Power Law + Peak} distribution or a PBH binary from the broken power law distribution. 

Under the above assumption, we performed hierarchical Bayesian analyses on the GWTC-3 data, and found that (1) PBHs described by a broken power law mass function are strongly preferred over the \textsc{Power Law + Peak} ABHs near $\sim 30 M_\odot$. These PBHs significantly suppress the ``peak'' in the \textsc{Power Law + Peak} model. (2) More than half of the GWTC-3 events can be attributed to PBHs. (3) Black holes with masses smaller than $30M_{\odot}$ are dominated by ABHs. (4) PBHs are rare below $m_{*}$, i.e., the mass function obeys a power law rather than a broken power law. One should note that these conclusions rely on the assumption that ABHs are described by the \textsc{Power Law + Peak} model, which is a phenomenological parametrization for the ABH mass function. Given the uncertainties in ABH formation, the strong preference over the \textsc{Power Law + Peak} ABH model is not conclusive evidence for the existence of PBHs.

These results impose several constraints and implications on our PBH mechanism. Firstly, the PBH mass function $f$ suggested by GWTC-3 is consistent with the prediction made in Ref.~\cite{Deng:2021ezy} for PBHs from supercritical bubbles: $f\propto m^{\alpha_2}$, where $\alpha_2 \sim -4$. Secondly, the best-fit value of the critical mass $m_{*}\sim30M_{\odot}$ leads to an estimate of the energy scale of the bubble interior, i.e., another vacuum that the inflationary state tunnels to during inflation: $\eta_{\text{b}}=\mathcal{O}(0.1)\ \text{GeV}$. Thirdly, almost all PBHs are from supercritical bubbles, which means most subcritical bubbles were destroyed or did not nucleate for some reason. This could happen if the quantum fluctuations on the bubble wall break the bubbles when they shrink. This gives the PBH a lower bound in mass. If we further assume that the energy scale of the bubble interior and that of the wall tension are comparable, we obtain a lower bound of the inflationary scale: $\eta_{i}>10^{4}\ \text{GeV}.$ Besides making inferences on the inflation model, we would like to emphasize that the observational evidence of the supercritical PBHs is also evidence of the multiverse. Moreover, the PBH mass distribution indicated by GWTC-3 can provide seeds of supermassive black holes located at the center of most galaxies~\cite{Serpico:2020ehh, Huang:2023chx}.

In this work, we assumed a constant nucleation rate of vacuum bubbles. However, the nucleation rate could vary with time~\cite{Huang:2023chx,Kleban:2023ugf}. In this case, the long wavelength primordial perturbations may affect the nucleation rate, leading to the initial clustering of the PBHs, which shall be studied in detail in future work. Moreover, when deriving the constraints on the inflation model, we assume that subcritical PBHs are suppressed due to the deviation from a spherically symmetric bubble. Actually, there are other reasons that may further suppress the subcritical bubbles. In particular, as the vacuum energy in our Universe changes during inflation, the surface tension of the bubble wall should also change accordingly. While in Ref.~\cite{Deng:2021ezy}, the surface tension has been treated as a free parameter so that the wall tension is not necessarily the same as that during inflation, the nonlinear effects, which have not been considered in Ref.~\cite{Deng:2021ezy}, can still lead to scalar waves peeling off from the wall, taking away additional energy. As a result, the subcritical bubbles might not collapse into black holes. This process might affect the inference on the inflation model, and will be investigated in more detail in future work with a full general relativity numerical simulation. In spite of the possible systemic errors caused by the uncertainties in modeling, this work shows a promising approach to probe the physics of the early universe with gravitational wave observations. This is especially the case with the next generation gravitational wave detectors, which are able to probe black hole mergers at high redshift.

\section{Acknowledgements}\label{sec:acknowledgements}
H. D. was supported by the U.S. Department of Energy, Office of High Energy Physics, under Award No.~de-sc0019470 at Arizona State University, and the National Science Foundation NANOGrav Physics Frontier Center No.~2020265. Y.-S. P. is supported by the National Natural Science Foundation of China (NSFC) under Grant No.~12075246 and the Fundamental Research Funds for the Central Universities. J. Z. is supported by the scientific research starting grants from the University of Chinese Academy of Sciences (Grant No.~118900M061), the Fundamental Research Funds for the Central Universities (Grant No.~E2EG6602X2 and Grant No.~E2ET0209X2), and the National Natural Science Foundation of China (NSFC) under Grant No.~12147103.


\bibliography{GWPBH}

\end{document}